\documentclass{llncs}
\include{amsthm}

\newcommand{\ceils}[1]{\lceil #1 \rceil}

\title{The Violation Heap: \\ A Relaxed Fibonacci-Like Heap}
\author{Amr Elmasry \thanks{Supported by an Alexander von Humboldt Fellowship.}\\
\institute{Max-Planck Institut f\"{u}r Informatik \\
Saarbr\"{u}cken, Germany \\
\small elmasry@mpi-inf.mpg.de}}

\date{}

\begin{document}
\maketitle 

\begin{abstract}
We give a priority queue that achieves the same amortized bounds as Fibonacci heaps. Namely, find-min requires $O(1)$ worst-case time, insert, meld and decrease-key require $O(1)$ amortized time, and delete-min requires $O(\log n)$ amortized time. Our structure is simple and promises an efficient practical behavior when compared to other known Fibonacci-like heaps. The main idea behind our construction is to propagate rank updates instead of performing cascaded cuts following a decrease-key operation, allowing for a relaxed structure.
\end{abstract}

\section{Introduction}

The binomial queue \cite{Vui} is a basic structure that supports the operations: find-min in $O(1)$ worst-case time, insert and meld in $O(1)$ amortized time, decrease-key and delete-min in $O(\log{n})$ worst-case time. It can also be extended to support insert in $O(1)$ worst-case time. Being so natural, simple and efficient, binomial queues do exist in most introductory textbooks for algorithms and data structures; see for example \cite{CLRS01}. 

Realizing that many important network-optimization and other algorithms can be efficiently implemented using a heap that better supports decrease-key, and that improving the bound for decrease-key is theoretically possible, Fredman and Tarjan \cite{FT87} introduced Fibonacci heaps supporting the operations: find-min in $O(1)$ worst-case time, insert, meld and decrease-key in $O(1)$ amortized time, and delete-min in $O(\log{n})$ amortized time. Using Fibonacci heaps the asymptotic time bounds for many algorithms have been improved; see \cite{FT87}. 

Around the same time, Fredman et al. introduced the pairing heaps \cite{FSST86}, a self-adjusting alternative to Fibonacci heaps. They only established $O(\log{n})$ amortized time bound for all operations. 
Stasko and Vitter \cite{SV87} improved the amortized bound for insert to $O(1)$. They also conducted experiments showing that pairing heaps are more efficient in practice than Fibonacci heaps and than other known heap structures, even for applications requiring many decrease-key operations! More experiments were also conducted \cite{MS94} illustrating the practical efficiency of pairing heaps. 
The bounds for the standard implementation were later improved by Iacono \cite{ia} to: $O(1)$ per insert, and zero cost per meld.
However, Fredman \cite{F99} showed that pairing heaps is not theoretically as efficient as Fibonacci heaps by giving a lower bound of $\Omega(\log\log{n})$, and precluded the possibility of achieving $O(1)$ decrease-key unless every node carries $\Omega(\log\log{n})$ information bits. Later, Pettie \cite{p} improved the analysis for the decrease-key operation to achieve $O(2^{2 \sqrt{\log \log n}})$ amortized bound. Recently, Elmasry \cite{Elm09} introduced a variant with $O(\log\log{n})$ amortized bound per decrease-key.  

Towards a heap that achieves good worst-case time bounds, Driscoll et al. \cite{DGST88} introduced the relaxed heaps. The rank-relaxed heaps achieve the same amortized bounds as Fibonacci heaps, and were easily extended to the run-relaxed heaps that achieve the bounds in the worst case except for meld. Relaxed heaps are good candidates for applications with possible parallel computations. Still, relaxed heaps are not practically efficient and are more complicated than Fibonacci heaps (they are not actually relaxed from this prospective). 
Another priority queue that achieves the same worst-case time bounds as run-relaxed heaps is the fat heap \cite{KST02}.
Incorporating $O(1)$ worst-case meld to the repertoire of operations, Brodal \cite{B96} introduced a priority queue that achieves the same bounds as the Fibonacci heaps, but in the worst-case sense. Brodal's structure is impractical and even more complicated than relaxed heaps.

Several attempts were made \cite{Hoy95,KT08,Pet87,Tak03} to come up with a priority queue that is theoretically as efficient as Fibonacci heaps without sacrificing the practicality. Among those, we find thin heaps \cite{KT08} the most natural and promising. In spite of being able to improve the space requirements by getting rid of the parent pointers \cite{KT08,Tak03}, or equivalently by using a binary-tree implementation \cite{Hoy95,Pet87}, the practicality issue is not resolved yet (or at least that is our impression!). 

In this paper we claim that we resolved this issue by introducing a priority queue that we call the \emph{violation heap}. Violation heaps have the same amortized bounds as Fibonacci heaps, and are expected to perform in practice in a more efficient manner than other Fibonacci-like heaps and compete with pairing heaps.
Our amortized bounds are: $O(1)$ per find-min, insert, meld and decrease-key, and $O(\log{n})$ per delete-min. 
In contrary to other Fibonacci-like heaps, while in agreement with pairing heaps, the \emph{degree} (number of children) of a node in the violation heaps is not necessarily logarithmic in the size of the subtree of this node; in fact there is no bound on node degrees, allowing what we call a \emph{relaxed structure}.
(Still, for the purpose of the analysis, the number of children of a node is amortized towards both the delete-min and decrease costs.) 

For Fibonacci heaps and thin heaps the degree of a node is bounded by $\approx 1.44 \lg{n}$, and for $2$-$3$ heaps \cite{Tak03} and thick heaps \cite{KT08} the bound is $\lg{n}$. Ensuring a larger degree bound (or even no bound) is not necessarily a disadvantage though; it only indicates how much relaxed the structure is. The reason is that a tighter degree bound may require more effort to restrict the structure to such bound. As an alibi, no such degree bound is guaranteed for pairing heaps \cite{FSST86}. Similar arguments can be mentioned about the bound on the height of a splay tree \cite{st} versus that of an AVL \cite{AVL} tree. In addition, one can resort to known techniques \cite{Elm04,EJK08} to reduce the number of comparisons performed in a delete-min operation almost by a factor of two.

In the next section we give our motivation: why there is a need for a new structure, what our objectives are, and how to achieve them. Then we introduce the data structure: design, operations and time bounds. Finally, we conclude the paper with some comments.

\section{Motivation}
In this section we argue why we need a new Fibonacci-like heap structure. We start with the drawbacks of other such heaps. Then, we summarize our objectives and the features required for a better heap structure. We end the section with ideas that lead to the violation heaps.

\subsection*{Drawbacks of other structures}

The pairing heap \cite{FSST86} is the most efficient among other Fibonacci-like heaps from the practical point of view \cite{MS94,SV87}. Still, it is theoretically inefficient according to the following fact \cite{F99}.

\begin{itemize}
\item The amortized cost per decrease-key is not a constant.
\end{itemize}

In contrary to pairing heaps, all known heaps achieving a constant amortized cost per decrease-key \cite{DGST88,FT87,Hoy95,KT08,Pet87,Tak03} impose the following constraint, which makes them practically inefficient.

\begin{itemize}
\item Every subtree has to permanently maintain some balancing constraint to ensure that its size is exponential with respect to its height.
\end{itemize}

\noindent The Fibonacci heaps \cite{FT87} have the following drawbacks.

\begin{itemize}

\item Every node has a parent pointer, adding up to four pointers per node. 

\item A subtree is cut and its rank decreases when its root loses two of its children. If the subtrees of these two children are small in size compared to the other children, the cascaded cut is immature. The reason is that the size of this cut-subtree is still exponential in its rank. This results in practical inefficiency as we are unnecessarily losing previously-gained information.

\item The worst-case cost per decrease-key can be $\Theta(n)$; see \cite[exercise 20.4-1]{CLRS01}.

\end{itemize}

Being able to remedy the drawbacks of Fibonacci heaps, other heap structures have not yet achieved the goal though! The trend of imposing more-restricted balancing constrains on every subtree would result in the following pitfalls that accompany a decrease-key operation \cite{Hoy95,KT08,Pet87,Tak03}.

\begin{itemize}
\item More cuts resulting in the loss of gained information. 
\item More checks among several cases resulting in performance slow down.
\end{itemize}

The rank-relaxed heaps \cite{DGST88} is even more restricted, allowing for no structural violations but for only a logarithmic number of heap-order violations. This requires even more case-based checks accompanying decrease-key operations.

\subsection*{Objectives for a new design}
To avoid the above drawbacks, we need a heap with the following properties.

\begin{itemize}
\item No parent pointers.
\item Structural violations in the subtrees of small children do not matter.
\item No cascaded cuts, and in accordance allowing a relaxed structure.
\item Fewer case-based checks following a decrease-key operation. 
\end{itemize}

\subsection*{Insights}
The main intuition behind our design is to allow structural violations resulting from decrease-key operations, and only record the amount of violations in every subtree within the root node of this subtree. 
We rely on the following three ideas: 

The first idea, which kills two birds by one stone, is to only consider violations in the first two children of a node (one child is not enough). As in \cite{KT08}, we utilize the unused left pointer of the last child to point to the parent. This makes it easy to convey such violations to a parent node without having a parent pointer.

The second idea, which we have recently used to improve the amortized cost of the decrease-key operation for pairing heaps to $O(\log \log n)$ \cite{Elm09}, is not to cut the whole subtree of a node whose key is decreased. Instead, we replace such subtree with the subtree of the last child. The expectation is that the size of a last-child's subtree constitutes a constant fraction of that of the parent, and hence the resulting structural degradation will be smoother. We also reduce the lost information resulting from the cut by keeping a good portion of the subtree. 

The third idea is about how to record the violations. As a compensation for the structural violation it has done, a decrease-key operation can pay $1/2$ credit to its parent, $1/4$ credit to its grandparent, and in general $1/2^i$ to its $i$-th ancestor. 
Once the credits in a node sum up to at least $1$, we declare its subtree as violating. Later, the fix of this violation can be charged to this credit. As long as the credits on a node are still less than $1$, its subtree is maintaining a good structure. The question is how to implement this idea! The details follow.

\section{The violation heaps}

\subsection*{Structure}

Similar to Fibonacci heaps, 2-3 heaps and thin heaps, the violation heap is a set of heap-ordered node-disjoint multiary trees. The children of a node are ordered according to the time when they are linked to the parent. Every node has three pointers and one integer in addition to its key, utilized as follows.

\renewcommand{\theenumi}{\alph{enumi}}
\begin{enumerate}
\item A singly-linked circular list of tree roots, with a root of minimum key first.
\item A doubly-linked list of children for each node, with a pointer to its last child.
\item For each last child, the unused pointer points to its parent.
\item An integer for each node representing its \emph{rank}. 
\end{enumerate}

We maintain the invariant that the size $s_z$ of the subtree of a node $z$ is exponential with respect to its rank $r_z$, but not with respect to its degree $d_z$. 
The \emph{violation} $v_z$ of a node $z$ indicates how bad the structure of its subtree is, and is defined in terms of $r_z$ and $d_z$ as 

\begin{eqnarray}
\nonumber
v_z = \left\{ 	\begin{array}{ll}
				d_z/2 - r_z & \mbox{~~~~if $d_z/2 - r_z > 0$,} \\
				0 & \mbox{~~~~otherwise.}
							\end{array}
			\right.
\end{eqnarray}
  
We emphasize that we only store $r_z$, but neither $d_z$ nor $v_z$. The notion of violations is only used in the analysis, but not in the actual implementation.

In the sequel, we call a node \emph{active} if it is one of the last two children of its parent.
We maintain the rank of a node $z$, in accordance with Lemma \ref{rank}, by updating $r_z$ once the rank of any of its active children $r_{z1}$ and $r_{z2}$ decreases. We use the following formula: 
\begin{equation}
\label{formula}
r_z \leftarrow \ceils{(r_{z1}+ r_{z2})/2} + 1.
\end{equation}
Evaluating this formula requires an integer addition, a right shift, and one or two increments. If $z$ has one or no children, the rank of a missing child is $-1$. 

The following primitive is used to consolidate the trees of the heap.

\begin{description}
\item
\emph{3-way-join}$(z,z1,z2)$ (the presumption is that $r_z = r_{z1} = r_{z2}$): \\
Assume w.l.o.g. that $z$'s value is not larger than that of $z1$ and $z2$. 
Ensure that the active child of $z$ with the larger rank is the last child. 
Make $z1$ and $z2$ the last two children of $z$ by linking both subtrees to $z$, and increment $r_z$. 
\end{description}

\subsection*{Operations}

\begin{itemize}

\item {\it find-min(h)}: Return the first root of $h$. \\

\item {\it insert(x,h)}: A single node $x$ is inserted into the root list of $h$. The rank of $x$ is initially set to zero. If the key of $x$ is smaller than the minimum of $h$, then $x$ is inserted in the first position, otherwise in the second position. \\

\item {\it meld($h_1,h_2$)}: The root lists of $h_1$ and $h_2$ are combined in a new list whose first root is the smaller between the minimums of the two heaps. \\

\item {\it decrease-key($\delta$,x,h)}: Subtract $\delta$ from the key of $x$. If $x$ is a root, stop after making it the first root if its new value is smaller than the minimum. If $x$ is an active node whose new value is not smaller than its parent, stop. 

Otherwise, cut the subtree of $x$ and glue in its position the subtree with the larger rank between its active children. 
Recalculate the rank of $x$ using (\ref{formula}).
Promote $x$'s subtree as a tree in $h$, and make $x$ the first root if its new value is smaller than the minimum.
Propagate rank updates by traversing the path of ancestors of $x$'s old position, as long as the visited node is active and as long as its recalculated rank using (\ref{formula}) is smaller than its old rank. \\

\item {\it delete-min(h)}: Remove from $h$ the first root and make each of its subtrees a tree in $h$. 
Repeatedly 3-way-join trees of equal rank until no three trees of the same rank remain. 
As for Fibonacci heaps \cite{FT87}, this is done in $O(1)$ time per tree using a temporary array indexed by rank values. 
Finally, the root with the new minimum value is moved to the first position in the root list.

\end{itemize}

\subsection*{Analysis}

First, we point out that 3-way-join is favorable to the normal join for our case.

\begin{lemma}
\label{3-way-join}
No extra violation units are added to the nodes of the heap as a result of a 3-way-join.
\end{lemma}  

\proof
Consider the 3-way-join$(z,z1,z2)$ operation, assuming that the value of $z$ is not larger than the values of $z1$ and $z2$.
When $z$ gains two extra children, $r_z$ is incremented ensuring that $d_z/2 - r_z$ does not change.
\qed

\bigskip

Next, we show that the assigned rank values fulfill the requirements.

\begin{lemma}
\label{rank}
The following relations are maintained for every node $z$.

\begin{eqnarray}
\nonumber
r_z \left\{ \begin{array}{ll}
        = 0 & \mbox{~~~if $z$ has no children,} \\
        \leq \ceils{(r_{z1} - 1)/2} + 1 & \mbox{~~~if $z$ has one child $z1$,} \\
        \leq \ceils{(r_{z1}+ r_{z2})/2} + 1 & \mbox{~~~if the active children of $z$ are $z1$ and $z2$.}
             \end{array}
      \right.      
\end{eqnarray}            
\end{lemma} 
 
\proof
When a node $z$ is inserted $r_z$ is set to $0$.
When a subtree of a node $x$ is cut by a decrease-key operation, the ranks of the ancestors of $x$ are updated by traversing the affected path upwards. As long as the violation of the nodes on the path is to be increased (rank should decrease), the decrease-key operation resumes the upward traversal and decreases the rank. Once the rank of a node is not to be changed, the rank of the other nodes along the path are already valid. For the case when the new rank is more than the old value no updates are done. 

When an active child vanishes, as a result of repeated decrease-key operations, another child (if there exist any) becomes active.   
This would make the inequalities given in the statement of the lemma satisfied as strict "$<$", and the recalculated rank is now larger than the rank stored in the parent. In such case, no rank updates are done and the upward traversal is terminated.

Consider the 3-way-join$(z,z1,z2)$, assuming that the value of $z$ is not larger than the values of $z1$ and $z2$. Before the join, $r_z = r_{z1} = r_{z2}$. After the join, $z1$ and $z2$ become the active children of $z$ implying that $r_z \leftarrow \ceils{(r_{z1} + r_{z2})/2} +1$. 
This explains the validity of incrementing $r_z$ after the join. 

If $z$ has one or no children, the lemma analogously follows by assuming the rank of a missing child to be $-1$.
\qed

\bigskip

The following two lemmas are used in the proof of Lemma \ref{structural} and Theorem \ref{main}.

\begin{lemma}
\label{tight}
Consider any node $z$. Let $z1$ and $z2$ be the active children of $z$, such that $r_{z1} \geq r_{z2}$. Then, either
\begin{enumerate}
\item $r_{z1} \geq r_{z}$, or
\item $r_{z1} = r_z - 1$ and $r_{z2} = r_z - 1$ or $r_z -2$.
\end{enumerate}
\end{lemma}

\proof
Using $r_z \leq \ceils{(r_{z1} + r_{z2})/2} + 1$ from Lemma \ref{rank}, then $r_z \leq r_{z1} +1$. Consider the case when $r_{z1} = r_z -1$. It follows that $r_z \leq \ceils{(r_z - 1 + r_{z2})/2} +1$, which implies $r_{z2} \geq r_z -2$. But $r_{z2} \leq r_{z1}$, indicating that $r_{z2}$ equals $r_z - 1$ or $r_z -2$.
\qed

\bigskip

\begin{lemma}
\label{critical}
The rank of any node can be decreased by at most $1$ when propagating rank updates within a decrease-key operation.
\end{lemma}

\proof
When a decrease-key is performed on a node $z$, the subtree of $z1$ is promoted in its position. From lemma \ref{tight}, $r_{z1} \geq r_{z} - 1$ implying that we now have a node that may be less in rank but by at most $1$, and the lemma holds. Propagating the rank updates using (\ref{formula}) would result in a decrease of $1$ in the rank of the parent node, if at all any.  
\qed

\bigskip

The above lemma is important and worth commenting. A crucial observation is that the parity of the ranks of the active children of a node affects the possibility of updating its rank. From Lemma \ref{rank}, if the sum of the ranks of the two active children is even and one of them is decremented, then the rank of the parent is not to be changed. If this sum is odd and one of the two ranks is decremented, then the rank of the parent may be decremented. Another consequence is that the sum of the violation units added to the heap nodes is bounded above by the number of rank-update steps performed within the decrease-key operations.

\bigskip

Now, we prove the {\em structural} lemma, illustrating that the size of a subtree is exponential with respect to its rank. 

\begin{lemma}
\label{structural}
Let $s_z$ be the size of the subtree of any node $z$, and $r_z$ be its rank. Then $s_z \geq F_{r_z}$, where $F_i$ is the $i$th Fibonacci number.
\end{lemma}

\proof
If $z$ has no children, then $r_z=0$ and the lemma holds. If $z$ has one child that is a leaf, then $r_z=1$ and $s_z=2$ and the lemma also holds. If $z$ has one child whose rank $r_{z1}>0$, then $r_z \leq \ceils{(r_{z1}-1)/2} +1$ implies $r_{z1} \geq r_z$. Using induction, $s_z > F_{r_{z1}} \geq F_{r_{z}}$. If $z$ has at least two children, while $r_{z1} \geq r_{z}$ the bound follows as above by induction.
Using Lemma \ref{tight}, the possibility left is that $r_{z1} = r_z - 1$ and $r_{z2} \geq r_z-2$.
Again using induction, then 
\[s_z > F_{r_{z1}} + F_{r_{z2}} \geq F_{r_z - 1} + F_{r_z - 2} = F_{r_{z}}.\]
\qed

\bigskip

\begin{corollary}
For any node $z$, $r_z = O(\log{s_z})$.
\end{corollary}

\bigskip

Finally, we prove the main theorem concerning the time bounds for the operations of the violation heaps.

\begin{theorem}
\label{main}
The violation heaps require $O(1)$ amortized cost per find-min, insert, meld, and decrease-key; and $O(\log{n})$ amortized cost per delete-min.
\end{theorem}

\proof

An active node is called \emph{critical} if the sum of the ranks of its active children is odd.
As a consequence of Lemma \ref{critical}, the decrease-key terminates the upward traversal for rank updates when it reaches 
\begin{enumerate}
\item a nonactive node, or
\item an active noncritical node (which becomes critical), or
\item a critical node whose rank is not to be changed (the current value is not larger than the recalculated value).
\end{enumerate}

Let $\gamma$ be the number of critical nodes in the heap. Let $\vartheta$ be the sum of violation units on the nodes of the heap, i.e. $\vartheta = \sum_{\forall z} (v_z = d_z/2 - r_z)$. Let $\tau$ be the number of trees in the heap. Let $\Delta_z$ be the number of decrease-key operations, since the 3-way-join in which $z$ was linked to its current parent, which have terminated at one of the two nonactive siblings of $z$ that were active just before the join. Let $\Delta = \sum_{\forall z} \Delta_z$.
We use the potential function
\[P = 3\gamma + 2\vartheta + \tau + \Delta.\] 

The actual time used by find-min, insert, and meld is $O(1)$.
Both find-min and meld do not change the potential, and an insertion increases $\tau$ by $1$. It follows that the amortized cost of these operations is $O(1)$. 

We call the path traversed by the decrease-key operation to perform rank updates the \emph{critical path}. The decrease-key uses $O(1)$ time plus the time it traverses the critical path.
The crucial idea is that every node on this path was critical and becomes noncritical after the traversal. In addition, the increase in violation units precisely equals the number of such nodes. Let $k$ be the number of nodes on a critical path. Then the actual time used by the decrease-key operation is $O(k)$. On the other hand, $\gamma$ decreases by at least $k-2$ (the cut node itself may also become a critical root), $\vartheta$ increases by $k$, $\tau$ increases by $1$, and $\Delta$ may increase by $2$ (in that case $\gamma$ decreases by $k-1$). Accordingly, the change in potential is at most $-3(k-2) + 2k + 1 = -k + 7$. These released $k$ credits pay for the $O(k)$ work done by the decrease-key, which only pays the $7$ credits.  

When the subtree of a node $z$ is cut and becomes a tree root after losing its last child $z1$ (which is glued in its position), the rank of $z$ is updated and may decrease, raising the need for more violation units. In such case, we claim that the increase in $2 \vartheta$ is at most the decrease in $\Delta$ plus $2$; these $2$ credits are also paid for by the decrease-key operation.
To prove our claim, consider the moment following a 3-way-join linking $z1$ to $z$. Let $r$ be the rank of $z1$, then the rank of $z$ is $r+1$. From Lemma \ref{tight}, the rank of its third-to-last child (whose rank is enforced to be larger than that of the fourth-to-last child) is at least $r-1$. When $z$ is cut and loses its last child $z1$, the third-to-last child of $z$ before the cut now becomes active with rank say $r'$. If $r' \geq r$, the rank of $z$ does not decrease and we are done. Otherwise, the decrease in the rank of $z$, which equals the increase in $\vartheta$, is at most $(r-r'+1)/2$. On the other hand, $\Delta_{z1}$ must have been at least $r-r'-1$ and is now reset to zero. Then, the increase in potential is at most $2(r-r'+1)/2 - (r-r'-1) = 2$, and the claim follows.
 
The preceding two paragraphs imply that the amortized cost for the decrease-key operation is $O(1)$. 

The 3-way-join uses $O(1)$ time, and decreases $\tau$ by $2$ resulting in a release of $2$ potential credits, which pay for the work. Note that the normal join may result in the root of the joined tree being critical, requiring more credits (that is the reason we resort to 3-way-join). 
In addition, from Lemma \ref{3-way-join}, no violations are added accompanying the 3-way-join.

Consider a delete-min operation for node $z$. When the children of $z$ become new trees, $\tau$ increases by $d_z$ and $\vartheta$ decreases by $v_z \geq d_z/2 - r_z$. Therefore, the change in potential is at most $2 r_z$, which is $O(\log{n})$ by Lemma \ref{structural}. Finding the new minimum requires traversing the roots surviving the consolidation, whose count is at most twice the distinct ranks because at most two roots per rank remain. It follows that the amortized cost for the delete-min operation is $O(\log n)$.  
\qed

\section{Comments}

We have given a priority queue that performs the same functions as a Fibonacci heap and with the same running-time asymptotic bounds. The main feature of our priority queue is that it allows any possible structure, with no structural restrictions soever, but only records the structure violations. 
Our priority queue uses three pointers and an integer per node. Following a decrease-key operation, we retain as much information as possible by not performing cascaded cuts and by keeping a big chunk of the subtree of the decreased node in its place. 
We expect our priority queue to be practically efficient. Experimental results still need to be conducted to support our intuitive claims. 

In comparison with other Fibonacci-like heaps, a drawback of violation heaps is that it cannot be implemented on a pointer machine. Indeed, we need the power of a RAM to recalculate the ranks using formula (\ref{formula}), which involves integer addition and bit operations.

Independently, and after the first technical-report version of this paper \cite{Elm08} (which is to-an-extent different from the current version) was archived, the idea of performing rank updates instead of cascaded cuts appeared in \cite{HST09}. 
The structure in \cite{HST09}, which is called rank-pairing heaps, relies on half-ordered half trees and not on multiary trees.
Also, the way the rank updating mechanism is performed in \cite{HST09} is different from that in violation heaps.

\end{document}